\documentclass[twocolumn,aps,prd,amsmath,amsfonts,nofootinbib,longbibliography]{revtex4-1}
\pdfoutput=1

\newcommand{\ud}{\mathrm{d}}

\newcommand{\bs}{\begin{split}}
\newcommand{\es}{\end{split}}

\usepackage{xcolor}
\usepackage{verbatim}
\usepackage{float}
\usepackage{color}
\usepackage[hyperindex,breaklinks]{hyperref}
\hypersetup{
colorlinks,
    linkcolor={black},
    citecolor={blue!50!black},
    urlcolor={blue!80!black},
    pdftitle={Path integrals with discarded degrees of freedom},
    pdfauthor={Luke M. Butcher}}
\usepackage{slashed}
\usepackage{graphicx}
\usepackage{subfigure}
\usepackage{psfrag}
\usepackage{mathtools}
\usepackage{amsthm}
\theoremstyle{plain}

\begin{document}

\title{Path integrals with discarded degrees of freedom}
\author{Luke M. Butcher}
\email[]{lmb@roe.ac.uk}
\affiliation{Institute for Astronomy, University of Edinburgh, Royal Observatory, Edinburgh EH9 3HJ, United Kingdom}
\date{February 14, 2019}

\begin{abstract}
Whenever variables $\phi=(\phi^1,\phi^2,\ldots)$ are discarded from a system, and the discarded information capacity $\mathcal{S}(x)$ depends on the value of an observable $x$, a quantum correction $\Delta V_\mathrm{eff}(x)$ appears in the effective potential \cite{Butcher18a}.  Here I examine the origins and implications of $\Delta V_\mathrm{eff}$ within the path integral, which I construct using Synge's world function. I show that the $\phi$ variables can be `integrated out' of the path integral, reducing the propagator to a sum of integrals over observable paths $x(t)$ alone. The phase of each path is  equal to the semiclassical action (divided by $\hbar$) including the same correction $\Delta V_\mathrm{eff}$ as previously derived. This generalises the prior results beyond the limits of the Schr\"odinger equation; in particular, it allows us to consider discarded variables with a history-dependent information capacity $\mathcal{S}=\mathcal{S}(x,\int^t f(x(t'))\ud t')$. History dependence does not alter the formula for $\Delta V_\mathrm{eff}$.
\end{abstract}

\maketitle
\section{Introduction}
In a recent paper \cite{Butcher18a} I obtained a powerful model for the quantum effect of discarded variables, applicable to any observable $x$ whose classical motion is determined by the nonrelativistic action
\begin{align}\label{redI}
\mathcal{I}[x(t)]\equiv\int \ud t \left[\frac{m}{2}\dot{x}^2 -V_\mathrm{cl}(x)\right].
\end{align}
In general, discarded variables $\phi\equiv(\phi^1,\ldots, \phi^d)$ with information capacity $\mathcal{S}(x)$ generate a quantum correction to the effective potential: $V_\mathrm{cl}\to V_\mathrm{cl} + \Delta V_\mathrm{eff}$, where
\begin{align}\label{VofS}
\!\Delta V_\mathrm{eff}= \frac{\hbar^2}{8m}\!\left[\!\left(\!1- 4\xi \frac{d+1}{d}\!\right)\!\!(\partial_x \mathcal{S})^2 + 2(1-4\xi)\partial_x^2 \mathcal{S}\right]\!,\!
\end{align}
for some $\xi \in \mathbb{R}$. The correction  (\ref{VofS}) appears in the Schr\"odinger equation for the wavefunction $\Psi_x(x,t)$ over the observable configuration space:
\begin{align}\label{reducedSchro}
i \hbar \partial_t \Psi_x = \left[-\frac{\hbar^2}{2m}\partial_x^2 + V_\mathrm{cl} +\Delta V_\mathrm{eff} \right]\Psi_x.
\end{align}
Consequently, $\Delta V_\mathrm{eff}$ directly affects the average motion of the observable:
\begin{align}\label{meanEOM}
m\partial_t^2\langle x\rangle&=- \langle \partial_x V_\mathrm{cl} + \partial_x\Delta V_\mathrm{eff}\rangle.
\end{align}
This motivates the use of a semiclassical action 
\begin{align}\label{semiclasS}
\mathcal{J}[x(t)]=\int \ud t \left[\frac{m}{2}\dot{x}^2 -\left(V_\mathrm{cl} + \Delta V_\mathrm{eff}\right)\right],
\end{align}
which generates trajectories consistent with the mean equation of motion (\ref{meanEOM}). This formalism allows us to model the quantum effect of discarded degrees of freedom when only their information capacity is known. For instance, when the holographic principle \cite{Bousso99b} is used to quantify the information capacity of the universe, the quantum correction (\ref{VofS}) provides a new explanation of cosmic acceleration \cite{Butcher18c}.

The results (\ref{VofS}--\ref{semiclasS}) were all obtained in the Schr{\"o}dinger picture. To complement that work, we now examine the origins and implications of $\Delta V_\mathrm{eff}$ in the path integral formalism. The paper is organised as follows. In section \ref{RedProps}, we show the quantum propagator can be written as an integral over observable paths $x(t)$ alone, with the phase of each path given by the semiclassical action (\ref{semiclasS}). In section \ref{PIcurved}, this same propagator is expressed as a path integral over the full configuration space, including the discarded variables. In section \ref{IntOut}, we reconcile these two viewpoints: by `integrating out' the paths $\phi(t)$, we generate the same quantum correction $\Delta V_\mathrm{eff}$ as the Schr\"odinger approach. Finally, in section \ref{NewS} we use the path integral method to extend the formula (\ref{VofS}) to cover \emph{history-dependent} capacities $\mathcal{S}=\mathcal{S}(x,\int^t f(x(t'))\ud t')$, which arise in the cosmological context when considering gauge transformations \cite{Butcher18c}. For ease of reference, key definitions from the previous paper \cite{Butcher18a} are included in the appendix.

\section{Propagator Over Observable Paths}\label{RedProps}
The main goal of this section is to construct the propagator $K(x_f,\phi_f,x_0,\phi_0;T)$ of a nonrelativistic particle over the full $(d+1)$-dimensional configuration space (\ref{genmetric}). We begin by observing that the reduced Schr\"odinger equation (\ref{reducedSchro}) can be solved with an ordinary one-dimensional path integral:
\begin{align}\nonumber
\Psi_x&(x_f,T)=\\\label{redprop}
&\int \ud x_0 \, \Psi_x(x_0,0)\int_{x(0)=x_0}^{x(T)=x_f} \mathcal{D}x(t)\, e^{i \mathcal{J}[x(t)]/\hbar},
\end{align}
where $\mathcal{J}[x(t)]$ is the semiclassical action (\ref{semiclasS}).

To understand the implications of equation (\ref{redprop}) we recal the definition (\ref{Psix}) of $\Psi_x$, and write an arbitrary state as
\begin{align}\label{genPsi}
\Psi &= \sum_{k} \frac{\Phi_k(\phi)}{[b(x)]^{d/2}}  \Psi^k_x(x,t),
\end{align}
where $\{\Phi_k\}$ are `energy' eigenfunctions (\ref{eigen}) with eigenvalues $\{E_\phi^k\}$, forming an orthonormal basis over the discarded configuration space $\mathcal{M}_\phi$:
\begin{align}\label{orth}
\int \ud^d\phi \sqrt{\tilde{g}(\phi)} \,\Phi_{k}(\phi) \Phi^*_{k'}(\phi) =\delta_{kk'}.
\end{align} 
As the Schr\"odinger equation (\ref{covSchro}) is linear, each $\Psi^k_x$ will obey its own reduced Schr\"odinger equation (\ref{reducedSchro}) with $E_\phi=E^k_\phi$ in the classical effective potential (\ref{Vcl}). Hence we can use equation (\ref{redprop}) to propagate each component of the general state (\ref{genPsi}):
\begin{align}\nonumber
&\Psi (x_f,\phi_f,T)=\sum_{k} \frac{\Phi_k(\phi_f)}{(b_f)^{d/2}}\\\label{PropComp}
&\quad \times\int \ud x_0\, \Psi^k_x(x_0,0)\int_{x(0)=x_0}^{x(T)=x_f} \mathcal{D}x(t)\, e^{i \mathcal{J}_k[x(t)]/\hbar},
\end{align}
where $b_f\equiv b(x_f)$  [similarly, we will write $b_0\equiv b(x_0)$, $\tilde{g}_f\equiv \tilde{g}(\phi_f)$, etc.] and the index on $\mathcal{J}_k[x(t)]$ indicates that we set $E_\phi=E^k_\phi$ in the classical effective potential.

We wish to represent the time evolution (\ref{PropComp}) as the result of a propagator $K(x_f,\phi_f,x_0,\phi_0;T)$ acting over the entire configuration space:
\begin{align}\label{Kdef}
&\Psi (x_f,\phi_f,T)\\ \nonumber
&=\int \ud x_0\, \ud^d \phi_0\,\sqrt{g_0}\, K(x_f,\phi_f,x_0,\phi_0;T)\Psi (x_0,\phi_0,0),
\end{align}
where  $\sqrt{g}=b^d \sqrt{\tilde{g}}$ is the covariant measure. It is easy to check that
\begin{align}\nonumber
&K(x_f,\phi_f,x_0,\phi_0;T)\\
\label{Ksum}
&=\sum_k \frac{\Phi_k(\phi_f)\Phi_k^*(\phi_0)}{(b_f b_0)^{d/2}}\int_{x(0)=x_0}^{x(T)=x_f} \mathcal{D}x(t)\, e^{i \mathcal{J}_k[x(t)]/\hbar}
\end{align}
is the propagator we need -- simply substitute (\ref{genPsi}) and (\ref{Ksum}) into the right of (\ref{Kdef}), perform the integral over $\phi_0$, and recover (\ref{PropComp}) as required. Equation (\ref{Ksum}) therefore achieves our stated goal: we have found an expression for the  propagator as a sum of integrals over observable paths $x(t)$ alone, with the phase of each path given by the semiclassical action (\ref{semiclasS}). The  prefactors $\Phi_k(\phi_f) \Phi^*_k(\phi_0)$ simply serve to project the wavefunction onto states of given $E_\phi$, so that the classical effective potential (\ref{Vcl}) can be evaluated within $\mathcal{J}[x(t)]$. 

Now, it must also be possible to represent the propagator as a path integral  over the \emph{entire} configuration space:
\begin{align}\label{Kpath}
K(x_f,\phi_f,x_0,\phi_0;T)=\int_{\substack{x(0)=x_0\\ \phi(0)=\phi_0}}^{\substack{x(T)=x_f\\ \phi(T)=\phi_f}} \mathcal{D}x(t)\,\mathcal{D}^d\phi(t)\,\ldots
\end{align}
How is this path integral related to the formula (\ref{Ksum}) we just derived? To answer this question, we must first construct the path integral (\ref{Kpath}) appropriate to our curved configuration space (\ref{genmetric}); this is covered in section \ref{PIcurved}. Then, in section \ref{IntOut}, we show that (\ref{Ksum}) follows directly from (\ref{Kpath}) by `integrating out' the paths $\phi(t)$. This calculation provides an independent derivation of $\mathcal{J}[x(t)]$ and $\Delta V_\mathrm{eff}$, establishing their validity beyond the Schr\"odinger picture.\footnote{Alternatively, we could start with a path integral over the entire \emph{phase space}: $\int \mathcal{D}x\,\mathcal{D}^d\phi\, \mathcal{D} p_x\,\mathcal{D}^d p_\phi \ldots$ Storchak has shown that this type of integral can be dealt with by transforming to a new time coordinate \cite{Storchak93} with $\Delta V_\mathrm{eff}$ then appearing as the Jacobian of the transformation. As this `canonical' approach favours a particular ordering of operators (rather than coordinate invariance) it tacitly sets $\xi=0$.} The path integral formalism therefore extends our results to domains where the Schr\"odinger equation no longer applies -- we shall explore this new territory in section \ref{NewS}.

\section{Path Integral Over Curved Space}\label{PIcurved}
Let us seek a general expression for the propagator of a nonrelativistic particle in a curved space of $D=d+1$ dimensions:
\begin{align}\label{propdef}
K(q_f,q_0;T)&\equiv\left\langle q_f \right|e^{-i\hat{H}T/\hbar}\left|q_0  \right\rangle,
\end{align}
where $q\equiv(q^1,q^2,\ldots,q^D)$ are arbitrary coordinates, and the Hamiltonian $\hat{H}$ is the operator in square brackets on the right of the covariant Schr\"odinger equation (\ref{covSchro}). The states $\{|q\rangle\}$ form a coordinate-invariant orthonormal basis,
\begin{align}\label{id}
\langle q'|q\rangle &= \frac{\delta^D(q'-q)}{\sqrt{g(q)}},&\int \ud^D q \sqrt{g(q)} \left|q\right\rangle \left\langle q \right|&=1,
\end{align}
and hence the propagator (\ref{propdef}) transforms as a scalar at both $q_f$ and $q_0$. 

To represent (\ref{propdef}) as a path integral, we split the exponential operator into $N$ equal factors, and insert $N-1$ copies of the identity (\ref{id}):
\begin{align}\label{skel}
K(q_f,q_0;T)&= \prod^{N-1}_{n=1}\int \ud^D  q_n \sqrt{g_n}\prod^N_{n=1} K(q_n,q_{n-1};\epsilon),
\end{align}
where $g_n\equiv g(q_n)$, $\epsilon \equiv T/N$, and $q_N\equiv q_f$. The path integral is then obtained as $N\to \infty$ and $\epsilon \to 0$.

\subsection{Short-Time Propagator}
To evaluate the path integral (\ref{skel}) we require a formula for the short-time propagator $K(q'\! ,q;\epsilon)$. To obtain it, let us begin with the following ansatz:
\begin{align}\nonumber
K(q'\! ,q;\epsilon)&= \sqrt{\frac{m}{2\pi i \hbar \epsilon}}^D\exp \left\{\frac{im }{\hbar \epsilon}\sigma (q'\! ,q)+ \alpha_0(q',q)\right.\\\label{Kans}
&\quad {} + \epsilon \alpha_1(q',q) +O(\epsilon^2)\Bigg\},\end{align}
where Synge's world function \cite{Synge}
\begin{align}\label{sigmadef}
\sigma(q',q)\equiv \left[\text{geodesic distance}(q', q)\right]^2\!/2
\end{align} 
provides a coordinate-invariant representation of the kinetic term. We will fix the unknown functions $\{\alpha_0,\alpha_1\}$ by insisting that the propagator (\ref{Kans}) satisfies the covariant Schr\"o\-ding\-er equation (\ref{covSchro}):
\begin{align}\label{KSchr}
i \hbar \partial_\epsilon K(q'\!,q;\epsilon)=\left[\frac{\hbar^2}{2m}\left(-\nabla^{2} + \xi R\right) + V_0\right]K(q'\!,q;\epsilon),
\end{align}
in the limit $\epsilon \to 0$. (Provided we are consistent, we can either evaluate $\{\nabla^2,R,V_0\}$ at $q$, or at $q'$; this choice will not affect the calculation.) Before doing so, let us first check that the propagator has the correct behaviour as $\epsilon\to 0$. Writing $\Delta q\equiv q'-q$, equation (\ref{sigmadef}) expands as
\begin{align}\label{sigmaexpand}
\sigma(q'\!,q)=g_{ij}\Delta q^i\Delta q^j/2+ O(\Delta q^{3}),
\end{align}
and so
\begin{align}\nonumber
\lim_{\epsilon\to 0}K(q'\!,q;\epsilon)&=\lim_{\epsilon\to 0}\sqrt{\frac{m}{2\pi i \hbar \epsilon}}^D\exp \left\{\frac{im}{2\hbar \epsilon}\bigg[g_{ij}\Delta q^i\Delta q^j  \right. \\ \nonumber
&\quad  {}  + O(\Delta q^{3}) \bigg] + \alpha_0(q',q)+ O(\epsilon) \bigg\}\\ \nonumber
&= \frac{\delta^D(\Delta q)}{\sqrt{\det (g_{ij})}}\exp \left\{\alpha_0(q,q)\right\}.
\end{align}
Hence the correct behaviour $K(q'\!,q;0)=\langle q' |q\rangle$ is achieved if and only if $\alpha_0(q,q)=0$. 

Returning to the covariant Schr\"o\-ding\-er equation (\ref{KSchr}), we insert 
the ansatz (\ref{Kans}) and sort the result into components proportional to $\epsilon^n K$: 
\begin{subequations}
\begin{align}\label{e-2}
\epsilon^{-2}K&: & 2 \sigma&= \nabla_i \sigma \nabla^i \sigma\\\label{e-1}
\epsilon^{-1}K&: &D&= \nabla^2 \sigma + 2 \nabla_i \sigma \nabla^i \alpha_0 \\ \nonumber
K&: &i\hbar \alpha_1&=\frac{\hbar^2}{2m}\bigg(-\nabla^2 \alpha_0   - \nabla_i \alpha_0 \nabla^i \alpha_0 \\\label{e0}
&&&\quad {} \left.{} -\frac{2im}{\hbar} \nabla_i \sigma \nabla^i \alpha_1 +\xi R\right)  + V_0.
\end{align}
\end{subequations}
Using the geometric identities \cite{Vines15},
\begin{align}\label{sigmaident1}
\nabla_i \sigma \nabla^i \sigma &= 2 \sigma, \qquad\quad  \left(\Rightarrow \nabla_i \sigma =O(\sigma^{1/2})\right)\\\label{sigmaident2}
\nabla^i \nabla_j \sigma &= \delta^i_j - \frac{1}{3}R^i{}_{kjl}\nabla^k \sigma \nabla^l \sigma+ O(\sigma^{3/2}),
\end{align}
we confirm that (\ref{e-2}) is automatically satisfied, and see that equation (\ref{e-1}) has solution
\begin{align}\label{a0sol}
\alpha_0 = \frac{1}{12}R^{ij}\nabla_i\sigma \nabla_j\sigma + O(\sigma^{3/2}).
\end{align}
This means $\alpha_0=O(\sigma)$ and hence $\alpha_0(q,q)=0$ as required; moreover, equation (\ref{e0}) becomes
\begin{align}\nonumber
i\hbar \alpha_1 &= \frac{\hbar^2}{2m}\left(- \frac{1}{6}R  - \frac{2im }{\hbar}\nabla_i\sigma \nabla^i \alpha_1+ \xi R \right)+V_0\\
&\quad + O(\sigma^{1/2}),
\end{align}
and is solved by
\begin{align}\label{a1sol}
\alpha_1 =- \frac{i}{\hbar}\left[\frac{\hbar^2}{2m}\left(\xi - \frac{1}{6}\right)R +V_0 \right] +O(\sigma^{1/2}).
\end{align}
Note that our solutions (\ref{a0sol}) and (\ref{a1sol}) can be evaluated with $\{g_{ij},\nabla_i,R_{ij},V_0\}$ at either $q$ or $q'$:  a change in convention  $q\leftrightarrow q'= q +O(\sigma^{1/2})$ will only generate terms at the neglected order.

Substituting (\ref{a0sol}) and (\ref{a1sol}) back into the ansatz (\ref{Kans}), we obtain
\begin{align}\nonumber
K(q'\! ,q;\epsilon)&= \sqrt{\frac{m}{2\pi i \hbar \epsilon}}^D\exp \left\{\frac{im}{\hbar \epsilon} \sigma (q'\! ,q) + \frac{ 1}{12}R^{ij}\nabla_i\sigma \nabla_j \sigma\right.\\ \nonumber
&\quad {} -\frac{i\epsilon}{\hbar}\left[\frac{\hbar^2}{2m}\left(\xi -\frac{1}{6}\right)R + V_0\right]\\\label{stprop}
&\quad {}+ O(\sigma^{3/2})+ O(\epsilon \sigma^{1/2}) + O(\epsilon^2)\bigg\}.
\end{align}
This is the short-time propagator we require, generating unitary evolution (\ref{propdef}) through $t\to t+\epsilon$ according to the covariant Schr\"odinger equation (\ref{covSchro}).

\subsection{Continuum Limit}
In general, terms $O(\epsilon^{3/2})$ do not contribute to the path integral (\ref{skel}) in the continuum limit: as $\epsilon \to 0$, we have $\prod^N_{n=1} \exp\{O(\epsilon^{3/2})\}=1+O(\epsilon^{1/2})\to1$. Moreover, due to rapid oscillations of the factor $\exp(im\sigma/\hbar \epsilon)$, the short-time propagator (\ref{stprop}) gives a negligible contribution to the integrals (\ref{skel}) for $\sigma \gg  \hbar\epsilon /m$. Hence we can treat $\sigma = O(\epsilon)$ and drop all the neglected terms in (\ref{stprop}). In a similar vein, note that
\begin{align}\label{manip}
0&=\left(\frac{2 \hbar \epsilon}{im}\right)\partial_{g_{ij}}\sqrt{\frac{2\pi i \hbar \epsilon}{m}}^D\\\nonumber
&=\left(\frac{2 \hbar \epsilon}{im}\right)\partial_{g_{ij}}\int \ud^D \Delta q \sqrt{g}\, e^{im g_{kl}\Delta q^k\Delta q^l/2\hbar\epsilon}\\ \nonumber
&=\int \ud^D \Delta q \sqrt{g}\, e^{im g_{kl}\Delta q^k\Delta q^l/2\hbar\epsilon}\!\left[\Delta q^i \Delta q^j +\left(\frac{2 \hbar \epsilon}{ im}\right)\frac{g^{ij}}{2}\right]\\ \nonumber
&=\int \ud^D q \sqrt{g}\, e^{im \sigma(q'\!,q)/\hbar\epsilon}\!\left[\nabla^i\sigma \nabla^j\sigma -\frac{i \hbar \epsilon }{m}g^{ij} + O(\epsilon^{3/2})\right]\!,
\end{align}
where equation (\ref{sigmaexpand}) was used in the last line. We can therefore set $\nabla_i\sigma \nabla_j \sigma= g_{ij}(i \hbar \epsilon /m)$ inside the path integral (\ref{skel}).

Applying the arguments of the previous paragraph to equation (\ref{stprop}) we arrive at a simplified version of the short-time propagator:
\begin{align}\nonumber
K(q'\! ,q;\epsilon)&= \sqrt{\frac{m}{2\pi i \hbar \epsilon}}^D\exp \left\{\frac{im }{\hbar \epsilon}\sigma (q'\! ,q) \right.\\\label{PIprop}
&\quad {}\left.{} -\frac{i\epsilon}{\hbar}\left[\frac{\hbar^2}{2m}\left(\xi -\frac{1}{3}\right)R + V_0\right]\right\},
\end{align}
suitable for the path integral (\ref{skel}) in the continuum limit. Comparing this formula with equation (10.153) of Kleinert's textbook \cite{Kleinert}, we confirm that our construction generalises the standard result to $\xi\ne0$.\footnote{Recal the path integral for a free particle in \emph{flat} space, composed of integrals over Cartesian coordinates $\int \ud^D x_n$ and propagators proportional to $\exp(im |x-x'|^2/2\hbar \epsilon)$. How would one generalise that expression to curved space, guided only by Einstein's equivalence principle and the notion of `minimal coupling'? It is natural to add the covariant measure $\sqrt{g(x_n)}$ to the integrals, and replace $|x-x'|^2\to 2 \sigma(x,x')$ as the invariant squared distance. However, there would be no reason to introduce $R$ to the propagator, as in equation (\ref{PIprop}). In other words, the minimally coupled path integral has $\xi =1/3$, not the value $\xi=0$ of the minimally coupled Schr\"odinger equation. We see that the meaning of `minimal coupling' depends on ones starting point. In general, this ambiguity justifies an agnostic view of curvature coupling: without a better argument to fix $\xi$, the parameter must be determined experimentally. \label{footmincoup}}

This completes our general treatment of the path integral over curved space. To construct the required path integral (\ref{Kpath}) we insert the short-time propagator (\ref{PIprop}) into equation (\ref{skel}), adopt the  metric (\ref{genmetric}), and take the limit $\epsilon \to 0$. 

\section{Integrating Out Discarded Paths}\label{IntOut}
Armed with a concrete formulation of the path integral over the full configuration space, we are now in a position to integrate out the discarded degrees of freedom $\phi$. 

\subsection{General method}
It will be useful to represent propagators in the $\{\Phi_k\}$ eigenbasis. Because $\hat{H}\Phi_k \propto \Phi_k$, the time evolution operator $\exp(-i\hat{H}t/\hbar)$ will not alter the $\phi$ dependence of any eigenfunction, so we must have
\begin{align}\label{Kexp}
K(x'\!,\phi'\!,x,\phi;t)= \sum_k \frac{\Phi_k(\phi')\Phi_k^*(\phi)}{\bar{b}^d}K_k(x'\!,x;t),
\end{align}
for some $K_k(x'\!,x;t)$. [The factors of $\bar{b}\equiv \sqrt{b(x')b(x)}$ are a convenient convention.] The components $K_k(x'\!,x;t)$ are then given by
\begin{align}\label{Kk}
&K_k(x'\!,x;t)\\ \nonumber
&=\bar{b}^d\int \ud^d\phi'  \ud^d \phi \sqrt{\tilde{g}(\phi')\tilde{g}(\phi)}\Phi^*_k(\phi')\Phi_k(\phi)K(x'\!,\phi'\!,x,\phi;t),
\end{align}
by virtue of orthonormality (\ref{orth}).

To see the value of this representation, let us set the coordinates $(x'\!,\phi'\!,x,\phi;t)\to(x_f,\phi_f,x_0,\phi_0;T)$ in equation (\ref{Kk}), insert the path integral (\ref{skel}) into the right hand side, and use the eigenbasis expansion (\ref{Kexp}) for the short-time propagators therein:
\begin{align}\nonumber
&K_k(x_f,x_0;T)\\ \nonumber
&=(b_fb_0)^{d/2}\int \ud^d\phi_f\,  \ud^d \phi_0 \sqrt{\tilde{g}_f\tilde{g}_0}\, \Phi^*_k(\phi_f)\Phi_k(\phi_0)\\ \nonumber
& \times\prod^{N-1}_{n=1}\int \ud x_n\, \ud^d \phi_n\, b_n^d\sqrt{\tilde{g}_n} \\
&\times \prod^N_{n=1}\sum_{k_n} \frac{\Phi_{k_n}(\phi_n)\Phi_{k_n}^*(\phi_{n-1})}{(b_n b_{n-1})^{d/2}}K_{k_n}(x_n,x_{n-1};\epsilon).
\end{align}
We can now perform the integrals over each $\phi_n$ in turn. The integral over $\phi_0$ leaves only the $k_1=k$ term in the sum over $k_1$, then the integral over $\phi_1$ leaves only the $k_2=k$ term in the sum over $k_2$, and so on. On completing the final integral (over $\phi_f\equiv \phi_N$) we have
\begin{align}\nonumber
&K_k(x_f,x_0;T) \\\nonumber
&=(b_fb_0)^{d/2} \prod^{N-1}_{n=1}\int \ud x_n\, b_n^d \prod^N_{n=1}\frac{K_{k}(x_n,x_{n-1};\epsilon)}{(b_n b_{n-1})^{d/2}}\\\label{Kkxpath}
&=\prod^{N-1}_{n=1}\int \ud x_n\, \prod^N_{n=1}K_{k}(x_n,x_{n-1};\epsilon),
\end{align}
which has the form of a path integral over  $x$ alone. 

Setting $(x'\!,\phi'\!,x,\phi;t)\to(x_f,\phi_f,x_0,\phi_0;T)$ in the eigenbasis expansion (\ref{Kexp}) and inserting the components (\ref{Kkxpath}) we arrive at
\begin{align}\label{Kskelsum}
&K(x_f,\phi_f,x_0,\phi_0;T)\\\nonumber
&=\sum_k \frac{\Phi_k(\phi_f)\Phi_k^*(\phi_0)}{(b_fb_0)^{d/2}}\prod^{N-1}_{n=1}\int \ud x_n\, \prod^N_{n=1}K_{k}(x_n,x_{n-1};\epsilon),
\end{align}
which is now very close to the formula (\ref{Ksum}) we wish to rederive. All that remains is to evaluate the short-time components $K_{k}(x_n,x_{n-1};\epsilon)$ and take the limit of equation (\ref{Kskelsum}) as $\epsilon \to 0$. 

\subsection{Calculation for $d=1$}
To illustrate the process described above, let us restrict our interest to the  configuration space (\ref{genmetric}) with $d=1$: 
\begin{align}\label{tubemetric}
\ud s^2 &= \ud x^2 + [b(x)]^2\ud \phi^2, & &\phi\in [0,2\pi),
\end{align}
which was the motivating example in the original paper \cite{Butcher18a}. In principle, the steps below can also be followed for any configuration space of the form (\ref{genmetric}).

The metric (\ref{tubemetric}) fixes $R= -2(\partial_x^2 b)/b$, $\tilde{g}=1$, $\tilde{R}=0$, and sets the eigenfunctions (\ref{eigen}) as 
\begin{align}\label{tubeEig}
\Phi_k=\frac{e^{i k\phi}}{\sqrt{2\pi}},\quad E_\phi^k=\frac{\hbar^2 k^2}{2m},\quad k\in \mathbb{Z}.
\end{align}
Consequently, when we insert the short-time propagator (\ref{PIprop}) into equation (\ref{Kk}) we get
\begin{align}\nonumber
K_k(x'\!,x;\epsilon)&= \frac{m \bar{b}}{2\pi i \hbar \epsilon}  \int \ud \Delta \phi\exp \left\{\frac{im }{\hbar \epsilon}\sigma (x'\!,x,\Delta \phi) - ik\Delta \phi \right.\\\label{shortk}
&\quad {}\left.{} -\frac{i\epsilon}{\hbar}\left[\frac{\hbar^2}{m}\left(\frac{1}{3}-\xi\right)\frac{\partial_x^2 b}{b} + V_0\right]\right\},
\end{align}
where $\Delta \phi \equiv \phi'-\phi$. To evaluate $\sigma(x'\!,x,\Delta \phi)$, we note that (\ref{sigmaident2}) implies
\begin{align}\nonumber
\partial_x^2\sigma &\to1, &\partial_\phi^2 \sigma&\to b^2,& \partial_\phi^2 \partial_x \sigma&\to b\partial_x b,\\
\partial_\phi^2 \partial_x^2 \sigma&\to2 b\partial_x^2 b /3,&  \partial_\phi^4 \sigma& \to - (b\partial_x b)^2,
\end{align}
as $x\to x'$ and $\Delta \phi \to0$, with all other $\partial^{(n)} \sigma\to0$ for $n\in \{0,1,2,3,4\}$. We can therefore construct a Taylor expansion in $\Delta x\equiv x'-x$ and $\Delta\phi$:
\begin{align}\nonumber
2\sigma (x + \Delta x,x,\Delta \phi) &= \Delta x^2 + \bar{b}^2 \Delta \phi^2 - \frac{b\partial_x^2 b}{6} \Delta x^2 \Delta \phi^2\\\label{sigmaTaylor}
&\quad {}- \frac{(b\partial_x b)^2}{12}\Delta \phi^4 +O(\sigma^{5/2}),
\end{align}
having used $\bar{b}^2= b[b + \Delta x\partial_x b+ \Delta x^2 \partial_x^2 b /2 + O(\sigma^{3/2})]$. 

We now substitute (\ref{sigmaTaylor}) into (\ref{shortk}) and perform the integral. In doing so, recall $\sigma = O (\epsilon)$ and that terms $     O(\epsilon^{3/2})$ can be neglected in the continuum limit; moreover, a previous trick (\ref{manip}) allows us to write $\Delta x^2= (i\hbar \epsilon/m) + O(\epsilon^{3/2})$. We therefore arrive at
\begin{align}\nonumber
K_k(x'\! ,x;\epsilon)&= \sqrt{\frac{m}{2\pi i \hbar \epsilon}} \exp \left\{\frac{i \epsilon}{\hbar}\left[\frac{m \Delta x^2}{2 \epsilon^2} \right.\right.\\ \label{KkEval}
&\quad {}\left.\left.{}-\left(V_0+ \frac{(\hbar k)^2}{2 m b^2}  + \Delta V_\mathrm{eff}\right)\right]\right\},
\end{align}
with $\Delta V_\mathrm{eff}$ given by equation (\ref{DeltaVD}) for $d=1$.

Note that $V_0 + (\hbar k)^2/2 m b^2$ is just the classical effective potential (\ref{Vcl}) with $E_\phi=E^k_\phi=(\hbar k)^2/2 m$ as prescribed by the eigenfunctions (\ref{tubeEig}). Thus equation (\ref{KkEval}) implies
\begin{align}\nonumber
&\prod^{N-1}_{n=1}\int \ud x_n\, \prod^N_{n=1}K_{k}(x_n,x_{n-1};\epsilon)\\\nonumber
&=\prod^{N-1}_{n=1}\int \ud x_n\, \prod^N_{n=1}\sqrt{\frac{m}{2\pi i \hbar \epsilon}} \exp \left\{\frac{i \epsilon}{\hbar}\left[\frac{m (x_n-x_{n-1})^2}{2 \epsilon^2} \right.\right.\\\nonumber
&\quad {}\left.\left.{}-\left(\left.V_\mathrm{cl}(x_n)\right|_{E_\phi=E^k_\phi} + \Delta V_\mathrm{eff}(x_n)\right)\right]\right\}\\
&\to  \int_{x(0)=x_0}^{x(T)=x_f} \mathcal{D}x(t)\, e^{i \mathcal{J}_k[x(t)]/\hbar}
\end{align}
as $\epsilon  \to 0$. Inserting this limit into (\ref{Kskelsum}) we finally recover the desired result (\ref{Ksum}).

\subsection{Remarks}
We have shown that the degrees of freedom $\phi$ can be integrated out of the path integral (\ref{Kpath}) over the curved configuration space (\ref{genmetric}). This renders the propagator as a sum of integrals over observable paths alone (\ref{Ksum}). Crucially, the phase of each observable path $x(t)$ is given by the semiclassical action (\ref{semiclasS}) which includes the same quantum correction (\ref{VofS}) as previously derived in the Schr\"odinger approach \cite{Butcher18a}.

Although the proof was only given explicitly for $d=1$, the same steps can be followed for any metric of the form (\ref{genmetric}). The result of this process must agree with (\ref{Ksum}) \emph{in general}, otherwise the path integral would be inconsistent with the covariant Schr\"odinger equation. Furthermore, when the Sch\"odinger equation cannot be used, the path integral still provides a route to obtain the quantum correction $\Delta V_\mathrm{eff}$ and the semiclassical action $\mathcal{J}[x(t)]$. This is illustrated in the next section.

\section{History-Dependent Information Capacity}\label{NewS}
Thus far, we have only discussed discarded variables whose information capacity depends on the present value of the observable: $\mathcal{S}=\mathcal{S}(x(t))$. For such systems, the full configuration space can be described by a curved metric (\ref{genmetric}) with a subspace $\mathcal{M}_{\phi}$ that changes size as a function of $x$. Consequently, the covariant Schr\"odinger equation (\ref{covSchro}) naturally determines the dynamics of the quantum state $\Psi(x,\phi,t)$.

But now suppose the discarded information capacity also depends on the \emph{history} of $x$:
\begin{align}\label{historyS}
\mathcal{S}&=\mathcal{S}(x,\eta[x(t)]), & \eta[x(t)] &\equiv \int_{0}^t f(x(t')) \ud t',
\end{align}
for some function $f: \mathbb{R}\to\mathbb{R}$. How should we model this system? Clearly, we cannot continue to use the curved space (\ref{genmetric}): this would require a radius $b=b(x,\eta[x(t)])$ that depends on the history of the particle, which cannot be defined in the Schr\"odinger approach. Fortunately, the path integral formalism provides a natural environment to quantify particle histories, and will accommodate the new form of information capacity (\ref{historyS}) without any great difficulty. To see how this works, we shall first consider the simplest nontrivial case $f=1$ (where the Schr\"odinger equation can still be used, with a minor modification) before moving on to the general case (\ref{historyS}).

\subsection{Time dependence}
For $f=1$, the information capacity (\ref{historyS}) reduces to
\begin{align}\label{St}
\mathcal{S}=\mathcal{S}(x(t),t).
\end{align}
This \emph{time-dependent} capacity can be represented in the Schr\"odinger picture by promoting $b(x)\to b(x,t)$ in the metric (\ref{genmetric}). However, because the configuration space is now time-dependent, the Schr\"odinger equation must become
\begin{align}\label{tSchro}
i \hbar \partial_t\!  \left(g^{1/4}\Psi\right) =\left[\frac{\hbar^2}{2m}\left(-\nabla^2 + \xi R\right) + V_0\right]\!\left(g^{1/4}\Psi\right),
\end{align}
in order that unitarity be preserved:
\begin{align}\nonumber
&\partial_t \int \ud ^D q\, \sqrt{g} |\Psi|^2 \\ \nonumber
&=\int \ud ^D q\,\left[  g^{1/4}\Psi^*\partial_t\! \left(g^{1/4}\Psi \right) + \partial_t \!\left(g^{1/4}\Psi^* \right)g^{1/4}\Psi\right]\\\nonumber
&=\frac{i\hbar}{2m}\int \ud ^D q\, \sqrt{g} \left[\Psi^*\nabla^2\Psi -\Psi\nabla^2\Psi^* \right]\\
&=0.
\end{align}
For the metric (\ref{genmetric}) with $b(x)\to b(x,t)$, the modified Schr\"odinger equation (\ref{tSchro}) is simply
\begin{align}\label{tSchroS}
i \hbar \partial_t\Psi =\left[\frac{\hbar^2}{2m}\left(-\nabla^2 + \xi R\right) + V_0 - \frac{ i \hbar d}{2}\partial_t \ln b\right]\Psi.
\end{align}
With this minor alteration, the derivation of $\Delta V_\mathrm{eff}$ then follows the same route as outlined in the appendix. Because the state (\ref{Psix}) contains a factor of $[b(x,t)]^{-d/2}$, the $i \hbar \partial_t$ operator on the left of the modified Schr\"odinger equation (\ref{tSchroS}) generates a term that exactly cancels the new term  $-(i \hbar d/2)\partial_t \ln b$ on the right. Consequently, the formula for the quantum correction (\ref{VofS}) is completely unchanged.

Turning our attention to the path integral, we see that time-dependent capacity (\ref{St}) modifies the formalism of sections \ref{PIcurved}--\ref{IntOut} in two ways. First, in order to solve the modified Schr\"odinger equation (\ref{tSchroS}) the short-time propagator (\ref{PIprop}) must now be
\begin{align}\nonumber
&K(q'\! ,q;t+\epsilon,t)\\\nonumber
&= \sqrt{\frac{m}{2\pi i \hbar \epsilon}}^D\exp \left\{\frac{im }{\hbar \epsilon}\sigma (q'\! ,q)  \right.\\\label{modProp}
&\quad {}\left.{} -\frac{i\epsilon}{\hbar}\left[\frac{\hbar^2}{2m}\left(\xi -\frac{1}{3}\right)R + V_0 - \frac{ i \hbar d}{2}\partial_t \ln b\right]\right\}.
\end{align}
Here, $\sigma(q'\!,q)$ is evaluated using the metric at $t$; hence when we assemble its Taylor expansion, as in equation (\ref{sigmaTaylor}), the factor of $\bar{b}^2$ will be replaced by $b(x'\!,t)b(x,t)$.

Second, the measures of the integrals (\ref{skel}) will now be $\sqrt{g_n}=(b_n)^d\sqrt{\tilde{g}_n}$, with
\begin{align}\label{bt}
b_n\equiv b(x_n,t_n)=b(x_n,n \epsilon).
\end{align}
In order that all the $b_n$ cancel when assembling the $x(t)$ path integral (\ref{Kkxpath}) we will need to define $\bar{b}\equiv \sqrt{b(x',t')b(x,t)}$ when constructing the components $K_k(x',x;t',t)$ as in (\ref{Kk}). Notice the difference between this replacement and the replacement for $\bar{b}$ in the Taylor expansion of $\sigma(q'\!,q)$; consequently, each integral over $\Delta\phi_n$ leaves a factor
\begin{align}\nonumber
&\left(\frac{b(x_n,t_n)}{b(x_n,t_{n-1})}\right)^{d/2}\\ \nonumber
&= \exp\left\{\frac{d}{2}\left[\ln b(x_n,t_n) -\ln b(x_n,t_n-\epsilon)\right]\right\}\\
&=\exp \left\{\frac{d}{2}\epsilon \partial_{t_n}\!\ln b_n+ O(\epsilon^{2})\right\},
\end{align}
by virtue of this mismatch. Fortunately, this factor is cancelled by the new term in the propagator (\ref{modProp}). Thus the time-dependence of the  information capacity (\ref{St}) has no effect on the results of our path integral calculation, besides making $\Delta V_\mathrm{eff}=\Delta V_\mathrm{eff}(x,t)$.

\subsection{History dependence}
We now turn to the history-dependent case (\ref{historyS}), which requires a radius $b= b(x,\eta[x(t)])$. For these systems, we cannot construct a Schr\"o\-dinger equation (\ref{tSchroS}) valid of  all $t\in [0,T]$ because $\eta[x(t)]= \int^t f(x(t'))\ud t'$ has no representation in terms of $\Psi (t)$. However, if we fix the time $t$, and \emph{assume} a specific history $x(t')\colon t'\in[0,t)$, then the radius $b= b(x,\eta[x(t)])$ returns to being a function of $x$ alone. Consequently, the Schr\"odinger equation (\ref{tSchroS}) is \emph{locally} well-defined, in that it allows us to evolve the wavefunction through a small interval $t\to t+ \epsilon$, for each possible history. This evolution is described by a short-time propagator (\ref{modProp}) with
\begin{align}\label{histpropterm}
\partial_t \ln b = \partial_t \ln b(x,\eta[x(t)])= f(x) \partial_\eta \ln b
\end{align}
therein. If we apply this propagator at $t=t_0,t_1,t_2\ldots $ in turn, and sum over every possible history, we naturally assemble the path integral. In this fashion, the finite propagator $K(x',x;T,0)$ can be constructed, from infinitesimal steps, without need for a \emph{global} Schr\"o\-dinger equation.

To actually perform the path integral, we note that the measures (\ref{bt}) become
\begin{align}
b_n&\equiv b(x_n,\eta_n),&\eta_n&\equiv \sum_{k=1}^{n} \epsilon f(x_k),
\end{align}
where the sum represents the integral that appears in equation (\ref{historyS}). Thus, exactly as above, each $\Delta\phi_n$ integral leaves us with a factor 
\begin{align}\nonumber
&\left(\frac{b(x_{n},\eta_{n})}{b(x_{n},\eta_{n-1})}\right)^{d/2}\\ \nonumber
&= \exp\left\{\frac{d}{2}\Big[\ln b(x_{n},\eta_{n}) -\ln b(x_{n},\eta_{n}-\epsilon f(x_n))\Big]\right\}\\
&=\exp \left\{\frac{d}{2}\epsilon f(x_n)\partial_{\eta_n}\! \ln b_n+ O(\epsilon^{2})\right\},
\end{align}
which cancels the new term (\ref{histpropterm}) in the short-time propagator (\ref{modProp}).

We therefore conclude that the formulae for the quantum correction (\ref{VofS}), the semiclassical action (\ref{semiclasS}), and the propagator (\ref{Ksum}) all remain valid  when the information capacity depends on the history of the observable (\ref{historyS}). Of course, as $\Delta V_\mathrm{eff}$ is now a function of the path $x(t)$, extra care must be taken when  actually evaluating the path integrals or generating the semiclassical equations of motion.

\section{Conclusion}
We have constructed the path integral (\ref{Kpath}) for a nonrelativistic particle living in the curved configuration space (\ref{genmetric}) and developed a general method by which the variables $\phi$ can be integrated out, and hence discarded. This procedure reduces the propagator to a sum of integrals over observable paths $x(t)$ alone (\ref{Ksum}). As the phase of each path is set by the semiclassical action (\ref{semiclasS}) this provides an independent derivation of the quantum correction $\Delta V_\mathrm{eff}$ previously obtained in the Schr\"odinger picture \cite{Butcher18a}. Thus, when the Schr\"odinger equation cannot be used (except over an infinitesimal time interval) the path integral grants us a means to calculate $\Delta V_\mathrm{eff}$ and hence model the quantum effect of discarded degrees of freedom. As an example of this generalisation, we have demonstrated that the formula for the quantum correction (\ref{VofS}) remains valid even when the discarded variables have information capacity that depends on the history of the observable (\ref{historyS}).

\begin{acknowledgements}
The author is supported by a research fellowship from the Royal Commission for the Exhibition of 1851, and the Institute for Astronomy at the University of Edinburgh.
\end{acknowledgements}

\appendix*

\section{Prior Results}
Here, we briefly recap the key definitions and results of the preceding work \cite{Butcher18a}. For detailed motivation and derivations, the reader should refer to the original paper.

Consider a nonrelativistic particle (of mass $m$) living on a curved $D=d+1$ dimensional space
\begin{align}\label{genmetric}
\ud s^2 &= g_{ij}\ud q^i\ud q^j= \ud x^2 + [b(x)]^2 \tilde{g}_{IJ}(\phi)\ud \phi^I \ud \phi^J,
\end{align}
on which there is also a potential $V_0(x)$. We wish to predict the behaviour of the `observable' coordinate $x$, while ignoring the variables $\phi\equiv(\phi^1,\ldots, \phi^d)$ that cover a $d$-dimensional compact manifold $\mathcal{M}_\phi$ with metric $\tilde{g}_{IJ}$ and physical volume $\mathrm{Vol}_\phi \propto b^d$. As usual, the metric $g_{ij}$ defines a covariant measure $\sqrt{g}\equiv \sqrt{\det(g_{ij})}$, a derivative operator $\nabla_i$, and curvature tensors $\{R^i{}_{jkl},R_{ij},R\}$ according to $[\nabla_i,\nabla_j]v^k= R^{k}{}_{lij}v^l$, $R_{ij}\equiv R^{k}{}_{ikj}$, $R\equiv R_{ij}g^{ij}$. Similarly, $\tilde{g}_{IJ}$ defines $\{\sqrt{\tilde{g}},\tilde{\nabla}_I,\tilde{R}^{I}{}_{JKL},\tilde{R}_{IJ},\tilde{R}\}$.

We can predict the $x$ coordinate of a \emph{classical} particle with the reduced action (\ref{redI}) where the effective potential
\begin{align}\label{Vcl}
V_\mathrm{cl}=V_0 + E_\phi/b^2
\end{align}
depends on the conserved `energy' 
\begin{align}\label{clasE}
 E_\phi\equiv \frac{1}{2m} \tilde{g}^{IJ}p_{\phi^I} p_{\phi^J}=\frac{mb^4}{2} \tilde{g}_{IJ}\dot{\phi}^I\dot{\phi}^J  =\mathrm{const}.
\end{align}
This allows us to treat the particle as though it were living on a reduced configuration space
\begin{align}\label{redmetric}
\ud s^2 &= \ud x^2,  & &x\in \mathbb{R},
\end{align}
without any reference to the variables $\phi$. $E_\phi$ is now viewed as a parameter of the system. 

For a \emph{quantum} particle, however, we must examine the behaviour of the wavefunction $\Psi(x,\phi,t)$, which defines probabilities via integrals of the form
\begin{align}\label{prob}
P = \int \ud^D q \sqrt{g}|\Psi|^2=\int \ud x \,\ud^d\phi \, b^d\sqrt{\tilde{g}}|\Psi|^2,
\end{align}
and obeys the covariant Schr\"odinger equation
\begin{align}\label{covSchro}
i \hbar \partial_t \Psi =\left[\frac{\hbar^2}{2m}\left(-\nabla^2 + \xi R\right) + V_0\right]\Psi,
\end{align}
for some $\xi\in\mathbb{R}$. (This free parameter reflects a quantisation ambiguity \cite{DeWitt57,DeWitt-Morette80,BLASZAK13}.  One might appeal to `minimal coupling'  and set $\xi=0$, however there is nothing particularly special about this choice -- see footnote \ref{footmincoup}.)

In order to separate the observable $x$ from the variables $\phi$, we consider states of the form 
\begin{align}\label{Psix}
\Psi = \frac{\Phi(\phi)}{[b(x)]^{d/2}}  \Psi_x(x,t),
\end{align}
where $\Phi$ is an eigenfunction over $\mathcal{M}_\phi$,
\begin{align}\label{eigen}
\frac{\hbar^2}{2m}\left(-\tilde{\nabla}^2  + \xi \tilde{R}\right)\Phi=E_\phi\Phi,
\end{align}
with unit norm
\begin{align}\label{Nphi}
\int \ud^d\phi \sqrt{\tilde{g}} |\Phi|^2=1.
\end{align} 
Note that $\Psi_x$ is normalised such that probabilities (\ref{prob}) become 
\begin{align}\label{redprob}
P = \int \ud x |\Psi_x|^2,
\end{align}
consistent with the metric on the reduced configuration space (\ref{redmetric}). Furthermore, the eigenvalue equation (\ref{eigen}) follows the same quantisation rule as the Schr\"odinger equation (\ref{covSchro}): $g^{ij} p_i p_j \to \hbar^2[-\nabla^2 +\xi R]$. 

Inserting (\ref{Psix}) into (\ref{covSchro}) we derive the reduced Schr\"o\-dinger equation (\ref{reducedSchro}) wherein the effective potential has a quantum correction
\begin{align}\nonumber
\Delta V_\mathrm{eff} &= \frac{\hbar^2 d}{2m} \left[ \left(\frac{d-2}{4} + \xi (1-d)\right)\left(\frac{\partial_x b}{b}\right)^2 \right.\\ \label{DeltaVD}
&\quad \left.{}+\frac{1-4\xi}{2}\left(\frac{\partial_x^2 b}{b}\right)\right].
\end{align}
To put this result in its final form (\ref{VofS}) we imagine dividing the curved space (\ref{genmetric}) into a lattice of small cells with spacing $\ell\ll\min\{ b,(b/\partial_x b),\sqrt{b/\partial_x^2 b}\}$. Then, at a given value of $x$, the information capacity of the $\phi$ subspace is 
\begin{align}\label{Sdef}
\mathcal{S}&= \ln \Omega = \ln (\mathrm{Vol}_\phi /\ell^d)= d \ln b - d \ln \ell + \mathrm{const}.
\end{align}
With this formula, it is easy to check that (\ref{VofS}) is equivalent to (\ref{DeltaVD}) for all $\ell>0$. Moreover, we are free to take the continuum limit $\ell \to 0$: the quantum correction (\ref{VofS}) remains finite even as $\mathcal{S}\to \infty$. 

\bibliography{Tube2}
\end{document}